\documentclass[prl,aps,twocolumn,superscriptaddress]{revtex4-2}
\usepackage{mathtools,bm,amsthm,amsmath,amsfonts,amssymb,graphicx,caption,subcaption,color,xcolor,appendix}
\usepackage[colorlinks,bookmarks=true,citecolor=blue,linkcolor=blue,urlcolor=blue,hyperfootnotes=false]{hyperref}

\usepackage[font=small,labelfont=bf]{caption}
\usepackage{footmisc}

\newcommand{\beqn}{\begin{eqnarray}}
\newcommand{\eeqn}{\end{eqnarray}}

\renewcommand{\bf}[1]{{\bm #1}}

\begin{document}

\title{Exact Parent Hamiltonians for All Landau Level States in a Half-flux Lattice}

\author{Xin Shen}
\thanks{These authors contributed equally to this work.}
\affiliation{School of Physics, Huazhong University of Science and Technology, Wuhan, 430074, China}

\author{Guangyue Ji}
\thanks{These authors contributed equally to this work.}
\affiliation{Department of Physics, Temple University, Philadelphia, Pennsylvania, 19122, USA}

\author{Jinjie Zhang}
\affiliation{Lamar Community College, Lamar, Colorado, 81052, USA}

\author{David E. Palomino}
\affiliation{Department of Physics, Temple University, Philadelphia, Pennsylvania, 19122, USA}

\author{Bruno Mera}
\affiliation{Instituto de Telecomunica\c{c}\~oes and Departmento de Matem\'{a}tica, Instituto Superior T\'ecnico, Universidade de Lisboa, Avenida Rovisco Pais 1, 1049-001 Lisboa, Portugal}
\affiliation{Advanced Institute for Materials Research (WPI-AIMR), Tohoku University, Sendai 980-8577, Japan}

\author{Tomoki Ozawa}
\affiliation{Advanced Institute for Materials Research (WPI-AIMR), Tohoku University, Sendai 980-8577, Japan}

\author{Jie Wang}
\thanks{X. Shen: shencongli.phys@gmail.com\\J. Wang: jie.wang0005@temple.edu}
\affiliation{Department of Physics, Temple University, Philadelphia, Pennsylvania, 19122, USA}

\begin{abstract}
    Realizing topological flat bands with tailored single-particle Hilbert spaces is a critical step toward exploring many-body phases, such as those featuring anyonic excitations. One prominent example is the Kapit-Mueller model, a variant of the Harper-Hofstadter model that stabilizes lattice analogs of the lowest Landau level states. The Kapit-Mueller model is constructed based on the Poisson summation rule, an exact lattice sum rule for coherent states. In this work, we consider higher Landau-level generalizations of the Poisson summation rule, from which we derive families of parent Hamiltonians on a half-flux lattice which have exact flat bands whose flatband wavefunctions are lattice version of higher Landau level states. Focusing on generic Bravais lattices with only translation and inversion symmetries, we discuss how these symmetries enforced gaplessness and singular points for odd Landau level series, and how to achieve fully gapped parent Hamiltonians by mixing even and odd series. Our model points to a large class of tight-binding models with suitable energetic and quantum geometries that are potentially useful for realizing non-Abelian fractionalized states when interactions are included. The model exhibits fast decay hopping amplitudes, making it potentially realizable with neutral atoms in optical lattices.
\end{abstract}

\maketitle

Narrow bands with desired topological and geometric properties are not only interesting themselves but also crucial for realizing exotic many-particle phases upon including interactions. Representative examples include the fractional quantum Hall effect in Landau levels (LLs)~\cite{RevModPhys.71.S298}. The ultra cold atom has been emerging as an important platform to design topological and many-body physics~\cite{RMP_Bloch_08,cooper:2019}. The Hofstadter topological bands are realized in optical lattices~\cite{aidelsburger:2013,miyake:2013}, and recently two-particle fractional quantum Hall state is also reported~\cite{Greiner_FQH_ColdAtom23}.

In searching for many-particle phases with fractionalization, interaction and single-particle band wavefunction both play a crucial role. The Kapit-Mueller (KM) model is one important step along this line in designing desired band properties, with experimental implications for electronic materials and neutral atoms in optical lattices~\cite{Kapit_Mueller}. The construction of the KM model relies on Poisson summation, an exact sum rule satisfied by lowest LL (LLL) states. As a result, KM model exhibits exact zero-energy flatband whose flatband Bloch states are nothing but LLL states sampled on lattice.

In this work, we consider the generalization of the KM model to stabilize higher LL states on a half flux lattice, which has the potential for hosting non-Abelian anyons when including interactions~\cite{MoreReadState,Read_Rezayi,RezayiHaldanePRL00}. Our construction starts from a generalization of Poisson summation from LLL to arbitrary high LLs~\cite{perelomov_completeness_1971,bargmann_completeness_1971,laughlin_spin_1989, Boon_Zak_81,Boon_Zak_83,abreu_discrete_2015}. Based on these new sum rules, we present explicit forms of tight-binding Hamiltonians $H_n$, with Gaussian-decaying hopping amplitudes, whose ground state is a zero-energy flatband, and the flatband Bloch states are $n_{\rm th}$ LL wavefunction sampled on a lattice which we term lattice LL states denoted by $\varphi_{n}$. We mainly focus on lattice with inversion and translation symmetry, and discuss features of our models: for odd LL index, the interplay between inversion symmetry and magnetic translation symmetry enforces $H_{2n+1}$ to be gapless and enforces $\varphi_{2n+1}$ to have singular points in the momentum space. Nevertheless, a large family of fully gapped tight-binding Hamiltonians can be constructed based on $H_{2n}$ and $H_{2n+1}$ which still preserves exact flatness of the ground states and offers a wide tunability of the flatband quantum geometry. Our work points out a designed construction of a family of tight-binding models that simultaneously have flat band and tunable band geometries, potentially realizable on optical lattices.

\emph{Review of the Kapit-Mueller model.---} We start with setting the conventions, followed by reviewing the standard LLL Poisson summation and the KM model~\cite{Kapit_Mueller}. Consider a two-dimensional infinite lattice generated by primitive vectors $\bm a_{1,2}$. The lattice is denoted as $\Lambda = \{m\bm a_1 + n\bm a_2\}$ where $m,n$ are integers, the area spanned by the unit cell is $|\bm a_1\times\bm a_2| = 2\pi l^2$. We denote the reciprocal lattice as $\bm b_{1,2}$, which obeys $\bm a_i\cdot\bm b_j = 2\pi\delta_{i,j}$. Additionally, we consider a continuum LL wavefunction, whose magnetic unit cell has area $2\pi l_B^2$, where $l_B$ is the magnetic length. So there is $\phi = l^2/l_B^2$ number of flux quanta per lattice unit cell. Lastly, we define lattice LL states as the continuum LL wavefunction sampled on lattice points, 
\begin{equation}
    |\varphi_{n}\rangle = \sum_{\bm r \in \Lambda} \Phi_{n}^{l_B}(\bm r)|\bm r\rangle,\label{deflatticeLL}
\end{equation}
where $|\bm r\rangle$ are local, orthonormal states obeying $\langle\bm r|\bm r'\rangle = \delta_{\bm r\bm r'}$. Coordinate $\bm r \in \Lambda$ is a lattice point. Here, $\Phi_n^{l_B}$ is the $n_{\rm th}$ LL wavefunction defined in continuum and $|\varphi_{n}\rangle$ is an un-normalized vector. Both $\Phi_{n}^{l_B}$ and $|\varphi_{n}\rangle$ possess good translation quantum numbers $\bm k$ on torus with enlarged unit cell containing integer number of flux quanta~\cite{haldaneholomorphic,Jie_MonteCarlo}.

In particular, the LLL space is spanned by holomorphic functions times the Gaussian factor,
\begin{equation}
    \Phi_{0}^{l_B}(\bm r) = f(z) \exp\left(-|z|^2/2l_B^2\right),
\end{equation}
where $z = w_a\bm r^a$ is the complex coordinate of $\bm r$ and $w_a$ is the complex structure. As most commonly used, we will take complex structure to be $w_{a} = (1,i)/\sqrt{2}$. In what follows, we use bold letters $\bm a_{1,2}$ and un-bold letters $a_{1,2}$ for lattice vectors in $\mathbb{R}_2$ and complex plane $\mathbb{C}_1$ respectively.
The LLL states are the coherent states of the LL ladder operator. Higher LL states are systematically constructed by applying the ladder operators $\hat a_{l_B}^\dag$ associated to magnetic length $l_B$, which is,
\begin{equation}
    \hat a_{l_B}^\dag = \bar z/l_B - l_B\partial_z,\quad \hat a_{l_B} = l_B\partial_{\bar z}.
\end{equation}
The operators above act only to the function $f(z)$ for LLL states but not to the Gaussian factor.

\begin{figure}[t]
\includegraphics[width=1\columnwidth]{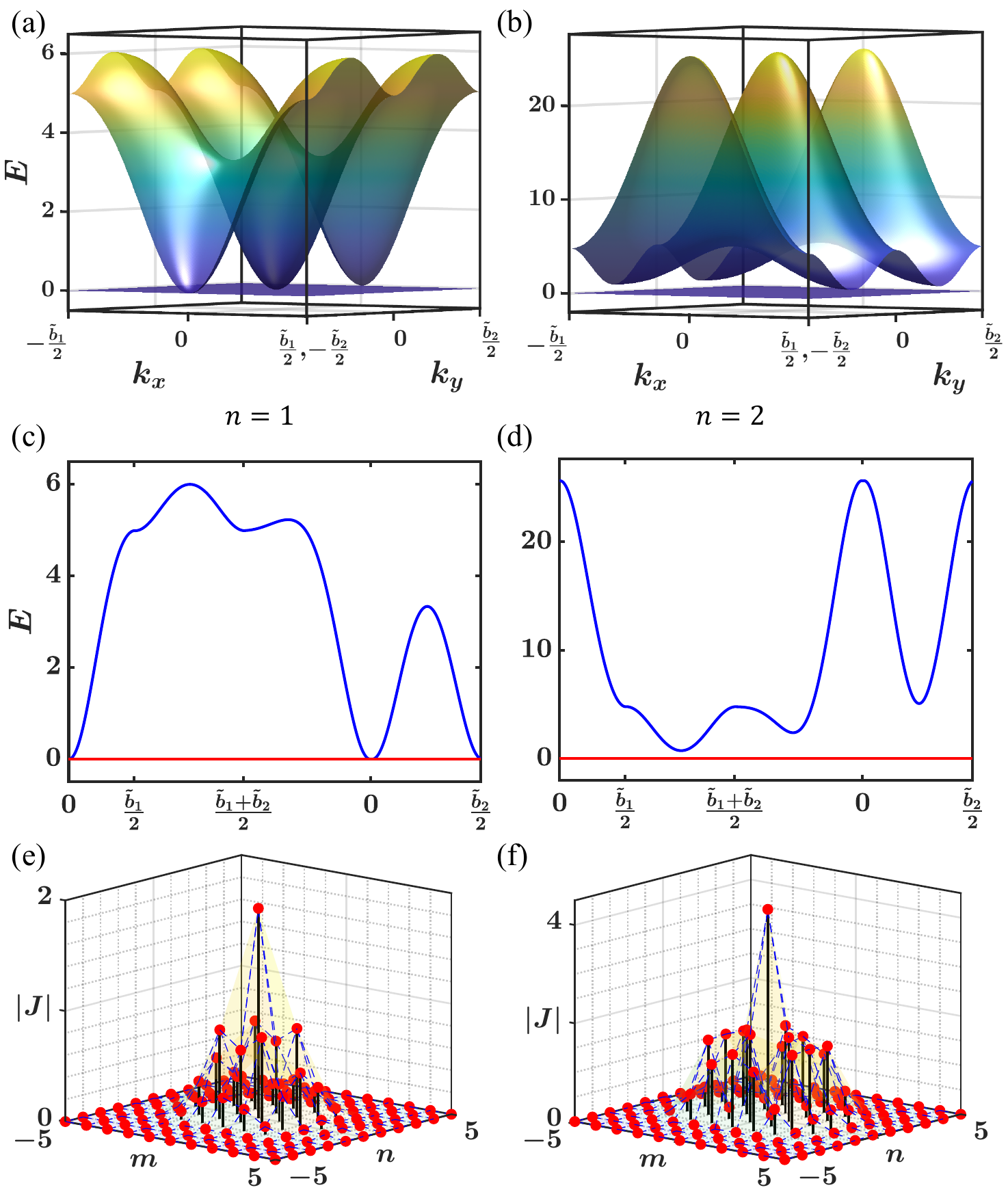}
\caption{\label{fig: energyband} Energy spectrum plotted under standard Landau gauge, and hopping amplitudes of $H_1$ and $H_2$ on rectangular lattice with aspect ratio $|\bm a_2| = 1.2 |\bm a_1|$. The $\tilde{\bm b}_{1}=\bm b_1/2$ and $\tilde{\bm b}_2 = \bm b_2$ are reciprocal vectors spanning the reduced Brillouin zone. (a), (b) are energy bands for $H_{1}$ and $H_{2}$, respectively. Their ground states form exact flat bands. The odd series $\hat H_{2n+1}$ have symmetry enforced quadratic band touching points. (c), (d) are the corresponding energy bands along high symmetric points. For $H_1$, it is gapless at high symmetry points $\bm{k}_{*}{=}(0,0)$ and $\tilde{\bm{b}}_2/2$ due to symmetries. Figure (e), (f) are the hopping amplitudes $\vert J(\bm d) \vert$ for $H_{1}$, $H_2$, respectively. The hoppings $J(\bm d)$ are defined in $H = \sum_{\bm r,\bm d\in\Lambda} J(\bm d) |\bm r-\bm d\rangle\langle\bm r|$. They have a concentrated distribution around the origin and decay fast away from the origin. The hopping amplitudes beyond NNN order are nearly negligible.}
\end{figure}

The KM model is a special type of Harper-Hofstadter model, whose hoppings are designed from Poisson summation rule~\cite{perelomov_completeness_1971,bargmann_completeness_1971,laughlin_spin_1989,Thomale_Poisson12} such that the model stabilizes lattice LLL state $|\varphi_0\rangle$ of any $\phi \in (0, 1)$ as zero-energy ground state. The Poisson summation dictates that any holomorphic function $f(z)$, or LLL states equivalently, are exactly annihilated by the following lattice summation,
\begin{equation}
    \sum_{\bm r\in\Lambda} \eta_{\bm r} f(z) e^{-\frac{|z|^2}{2l^2}} = \sum_{\bm r\in\Lambda} \eta_{\bm r} \Phi_0^{l_B = l}(\bm r) = 0,\label{PoissonLLL}
\end{equation}
as long as $f(z)$ times Gaussian is not diverging. The gauge function $\eta_{\bm r} = (-1)^{m+n+mn}$ is defined respect to lattice points $\bm r = m\bm a_1 + n\bm a_2$. Rewriting Eqn.~(\ref{PoissonLLL}) yields a lattice sum rule applicable for LLL states of length scale $l_B \neq l$,
\begin{equation}
    \sum_{\bm r\in\Lambda} \eta_{\bm r} e^{-\frac{|z|^2}{2}(l^{-2} - l_B^{-2})}\Phi^{l_B}_{0}(\bm r) = 0.\label{PoissonLLL2}
\end{equation}
When $l < l_B$, the above sum rule motivates the construction of the KM Hamiltonian $\hat D_0$,
\begin{eqnarray}
    \hat D_0 &=& \sum_{\bm d \in \Lambda} W_0(\bm d) \hat t(\bm d),\\
    W_0(\bm d) &=& \eta_{\bm d} e^{-\frac{|d|^2}{2}(l^{-2} - l_B^{-2})},
\end{eqnarray}
where $\hat t(\bm d)$ is the magnetic translation operator defined on lattice satisfying,
\begin{equation}
    \hat t(\bm d)\hat t(\bm d') = \exp\left(-i\bm d\times\bm d'/l_B^2\right)\hat t(\bm d')\hat t(\bm d),
\end{equation}
for lattice points $\bm d, \bm d' \in \Lambda$. See appendix for expressions of the standard lattice magnetic translation operator $\hat t(\bm d)$. The KM Hamiltonian $\hat D_0$ is hermitian and non-negative, and annihilates the lattice LLL states: $\hat D_0 |\varphi_0\rangle = 0$~\footnote{Comparing the original Kapit-Mueller model~\cite{Kapit_Mueller}, we have included an on-site chemical potential term such that the lattice LLL states are realized as zero modes.}.

\emph{Higher LL sum rule and exact parent Hamiltonians.---} We proceed to discuss the key results of this work, including lattice sum rules for higher LL states and the corresponding exact parent Hamiltonians, on a half-flux lattice. Firstly we notice that, any even-index LL states with $l_B = l$ are exactly annihilated when being summed on lattice weighted by the gauge function,
\begin{equation}
    \sum_{\bm r\in\Lambda} \eta_{\bm r} \Phi^l_{2n}(\bm r) = 0.\label{Poisson2nLL}
\end{equation}
Eqn.~(\ref{Poisson2nLL}) can be viewed as the generalized Poisson sum rule. It follows from replacing the LLL in Eqn.~(\ref{PoissonLLL}) by a squeezed coherent state
--- a LLL state of a different complex structure.
Since squeezed coherent state only has even LL weights, projecting it into a particular LL, one obtains Eqn.~(\ref{Poisson2nLL}).
In fact, Eqn.~(\ref{Poisson2nLL}) was initially derived by M. Boon and J. Zak~\cite{Boon_Zak_81,Boon_Zak_83} when studying amplitudes on von Neumann lattices. It is worth mentioning that Eqn.~(\ref{Poisson2nLL}) is valid only for even index LLs and requires $l_B = l$; it does not apply for odd LLs, and the non-decaying weight in the sum rule does not leads to parent Hamiltonians. Nevertheless, in what follows, we utilize Eqn.~(\ref{Poisson2nLL}) to derive sum rules for all LL states on a half flux lattice, including the odd-index LLs, where the fractional flux condition $\phi = 1/2 < 1$ makes the construction of parent Hamiltonians possible.

To motivate the construction of parent Hamiltonians for higher lattice LL states, we examine Eqn.~(\ref{Poisson2nLL}) for second LL $\Phi_{n=2}$. It is expressed in terms of LL ladder operators as,
\begin{equation}
    \sum_{\bm r\in\Lambda} \eta_{\bm r} \left[(\bar z - l^2\partial_z)^2f(z)\right] e^{-\frac{|z|^2}{2l^2}} = 0.
\end{equation}
Applying LLL summation Eqn.~(\ref{PoissonLLL}) to holomorphic function $\partial_z^2f(z)$ and introducing a new Gaussian decaying factor, the above is rewritten in below as,
\begin{equation}
    \sum_{\bm r\in\Lambda} \eta_{\bm r} \bar z e^{-\frac{|z|^2}{2}(l^{-2} - l_B^{-2})} \left[(\bar z/l - 2l\partial_z)f(z)\right]e^{-\frac{|z|^2}{2l_B^2}} = 0.
\end{equation}

It is crucial to notice that $\bar z/l - 2l\partial_z$ is trivially related to the LL raising ladder operator $\hat a^{\dag}_{l_B}$, provided flux per unit cell is one half $l^2_B = 2 l^2$. Therefore we have shown that generic first LL states $[\hat a^{\dag}_{l_B}f(z)]\exp\left(-|z|^2/2l_B^2\right)$ also admit a sum rule. One can carry out similar procedures to prove lattice sum rules for all LL states in a half-flux lattice,
\begin{gather}
    0 = \sum_{\bm r\in\Lambda} W_n(\bm r) \Phi_n^{l_B}(\bm r),\quad \quad l_B = \sqrt2l,\\
    W_n(\bm d) = \eta_{\bm d} \bar{d}^n e^{-\frac{|d|^2}{2}(l^{-2} - l_B^{-2})},\label{defWn}
\end{gather}
whose associated zero-mode operator, satisfying $\hat D_n|\varphi_n\rangle = 0$, is given in Eqn.~(\ref{defDn}),
\begin{equation}
    \hat D_n = \sum_{\bm d\in\Lambda} W_n(\bm d) \hat t(\bm d).\label{defDn}
\end{equation}

Note $\hat D_{n>1}$ is not Hermitian. Nevertheless, a semi-positive definite hermitian Hamiltonian can be constructed for all LL indices $\hat H_n = \hat D^\dag_n \hat D_n$, which are the parent Hamiltonians proposed in this work. Because $\hat H_n$ has non-negative eigenvalues and $\hat D_n|\varphi_n\rangle = 0$, the exact zero-energy flatband $|\varphi_{n\bm k}\rangle$ will be stabilized as the ground state of $\hat H_n$.

We find that, while the flat band of the even index $\hat{H}_{2n}$ is generally gapped, the odd-index $\hat H_{2n+1}$ has quadratic band touching points. The gapless nature of $\hat H_{2n+1}$ is enforced by the inversion and magnetic translation symmetries. See the band structure of $\hat H_1$ and $\hat H_2$ in Fig.~\ref{fig: energyband} (a,b) for an illustration. This work focuses on generic Bravais lattices with only inversion and translation symmetries. Other spatial symmetries, such as four and six fold rotations on square and triangular lattices, can result in additional gapless points for certain $H_n$. We leave detailed considerations on this for future work.

In what follows, we first explain the origin of gapless points from two complementary views: operators and wavefunctions. Next, we discuss how to break symmetry constraints to open the gap while preserving the exact flatband. In the end, the effect of truncating the tight-binding hopping terms, possible many-body phases, and implementation in ultracold platforms are addressed.

\emph{Inversion and translation enforced quadratic band touching and singular points in odd series.---}
We define the inversion as the unitary transformation that maps the local basis $|\bm r\rangle$ to $|{-}\bm r\rangle$. It is straightforward to see that, the zero-mode operator transforms under inversion as $\hat D_n \rightarrow (-1)^n\hat D_n$. This ensures that, at inversion symmetric momentum points $\bm k_*$, the zero-mode operator of odd index will vanish identically because of $\hat D_{n,\bm k_*} = (-1)^n \hat D_{n,\bm k_*}$. As a result, the Hamiltonian becomes a two by two zero matrix at $\bm k_*$, leading to the closure of the spectrum gap. There are four inversion enforced gapless points $\bm 0$, $\bm b_1/2$, $\bm b_2/2$ and $(\bm b_1+\bm b_2)/2$, and two of them $\bm 0$ and $\tilde{\bm b}_2/2$ appear in the reduced Brillouin zone spanned by $\tilde{\bm b}_1 = \bm b_1/2$, $\tilde{\bm b}_2 = \bm b_2$ where band structure is plotted Fig.~\ref{fig: energyband}.

Interestingly, inversion and magnetic translation symmetries also enforce singular points for odd lattice LL states $\varphi_{2n+1}$ at inversion symmetric points $\bm k_*$. The singular points are defined as the momentum points where $\varphi_{2n+1}$ vanishes identically and cannot be removed by multiplying an overall function of momentum~\cite{rhim:yang:19}. Technically, this means that the corresponding orthogonal projector is not defined at that point and the map from the Brillouin zone to the projective space ceases to be well-defined at that point. To better illustrate the singular points in this two-band system, we rewrite the zero-mode lattice LL state as a two-component spinor. Its explicit form is provided as follows, derived using the magnetic translation symmetry of continuous LL states,
\begin{equation}
    |\varphi_{n\bm k}\rangle = \Phi_{n\bm k}(\bm 0)|R_{\bm k}\rangle + \Phi_{n\bm k}(\bm a_1)|R'_{\bm k}\rangle,
\end{equation}
where the magnetic unit cell is chosen to be spanned by $2\bm a_1$ and $\bm a_2$, $\bm k$ is the associated translation momentum. In the above, $|R_{\bm k}\rangle$ and $|R'_{\bm k}\rangle$ are orthogonal states. $|\varphi_{n\bm k}\rangle$ is hence essentially a pseudo-spin-$1/2$ state, providing a mapping from the Brillouin zone to the Bloch sphere. Due to the parity of LL state, both $\Phi_{n\bm k}(\bm 0)$ and $\Phi_{n\bm k}(\bm a_1)$ vanish at inversion symmetric points $\bm k_*$ when $n$ is an odd integer. See appendix for details.

It is useful for later purposes to discuss the singular value decomposition (SVD) of the zero-mode operators. Since $\hat D_n$ commutes with the magnetic translation operator, it can be block-diagonalized by the momentum quantum number. At given momentum, $\hat D_{n\bm k}$ is a two by two non-hermitian matrix when $n \geq 1$. It admits a singular value decomposition,
\begin{equation}
    \hat D_{n\bm k} = \lambda_{n\bm k} |\varphi_{0\bm k}^{\perp} \rangle\langle \varphi_{n\bm k}^{\perp}|,
\end{equation}
where $\lambda_{n\bm k}$ is the non-zero singular value, and $\varphi_{n\bm k}^{\perp}$ is defined to be a normalized vector that is orthogonal to $\varphi_{n\bm k}$. Since we are working in a two band system, $\varphi_{n\bm k}^{\perp}$ is uniquely determined, up to a $U(1)$ phase, by $\varphi_{n\bm k}$ as long as it is defined --- this is generally true for all $\hat D_{2n,\bm k}$, and $\hat D_{2n+1,\bm k}$ when away from gapless points $\bm k_*$. Hence the SVD representation is restricted for even-index zero-mode operators, or odd indexed zero-mode operators outside gapless points. In the above, the left and right singular vectors are determined from the zero-mode equations $\hat D_n |\varphi_{n\bm k}\rangle = \hat D^\dag_n |\varphi_{0\bm k}\rangle = 0$~\footnote{Here the second zero-mode equation $\hat D^\dag_n|\varphi_0\rangle = 0$ follows directly from the LLL sum rule Eqn.~(\ref{PoissonLLL2}).}.

The SVD representation implies $\hat H_n$ and $\hat D_n\hat D^\dag_n$ share identical spectrum: one zero-mode and one dispersive band $E_{n\bm k} = |\lambda_{n\bm k}|^2$, but their eigenstates are different. Suppose $\lambda_{n\bm k}$ is smooth with respect to $\bm k$, the quadratic form of Bloch energy $E_{n\bm k}$ also immediately implies its vanishing behavior cannot be linear but has to be quadratic. Last but not least, the SVD representation implies one way to construct gapped parent Hamiltonians, to be explained in below.

\begin{figure}[t]
\includegraphics[width=1\columnwidth]{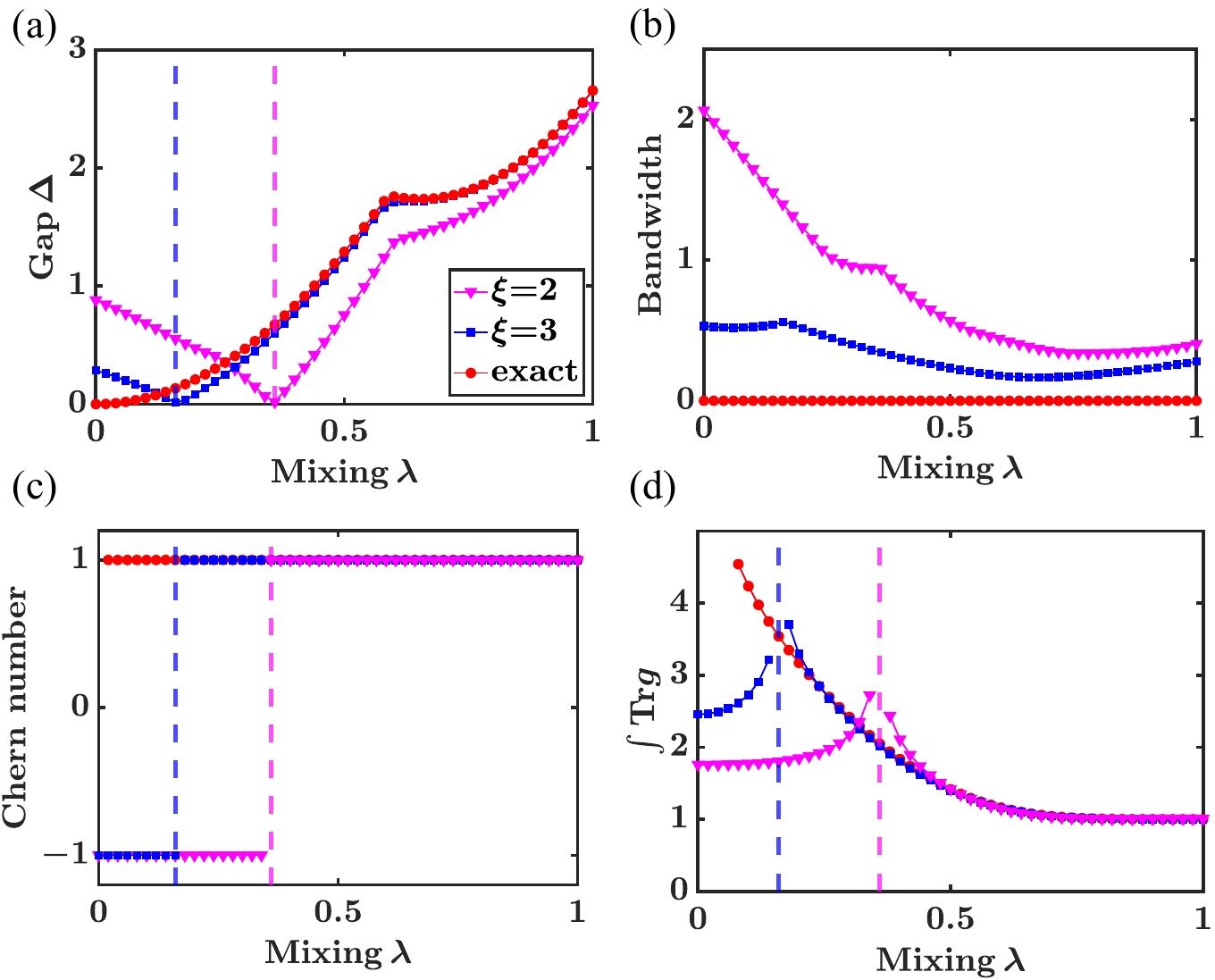}
\caption{\label{fig: mix_and_truncated} The evolution of band properties with respect to the mixing $\lambda$ and the hopping truncation $\xi$. The truncation sets the upper bound of the hopping range such that we only retain hopping terms $J(m\bm a_1+n\bm a_2)$  with $\vert m \vert {+} \vert n \vert {\le} \xi$ in model Eqn.~(\ref{def_lambda_model}). (a) Band gap, which is defined as the minimal energy of the higher band minus the maximal energy of the lower band. (b) Bandwidth of the lower band. (c), (d) Chern number and the integration of the trace of metric tensor. While the exact model with $\xi = \infty$ is fully gapped with a $C = 1$ lower band whenever $\lambda \neq 0$, the truncated model with finite $\xi$ has topological transition at finite $\lambda_c$ (marked with dash lines) where gap closes and reopens. The critical $\lambda_c$ moves towards $0$ when $\xi$ increases.
}
\end{figure}

\emph{Gapped exact parent Hamiltonians.---}
Here we consider linearly mixing $\hat D_n$ by complex valued superposition coefficients $\kappa_n\in\mathbb{C}$,
\begin{equation}
    \hat H_\kappa \equiv \hat D^\dag_\kappa \hat D_\kappa,\quad \hat D_\kappa \equiv \sum_n \kappa_n \hat D_n.
\end{equation}

The SVD form of $\hat D_n$ implies $\hat H_\kappa$ still has one exact zero-energy ground state whose wavefunction is the vector orthogonal to $\sum_n \bar\kappa_n \lambda_{n\bm k}|\varphi_{n\bm k}^\perp\rangle$. Importantly, when $\{\kappa_n\}$ contains both even and odd components, due to the loss of inversion constrains, $\hat H_\kappa$ becomes gapped.

Numerically, first LL like topological flatband obeying $\int{\rm Tr}g{\approx}3$ is found to be crucial for interacting fermions to form non-Abelian states, supported by various settings including standard LLs, moiré flatbands~\cite{DiXiaoNonabelian24,YangZhangNonabelian24,ChoNonabelian24,LiangFuNonabelian24,Fujimoto24} and the general framework termed generalized LLs~\cite{GLL_JieWang_2024}. Motivated by this, we examine first LL type gapped flatband in our model by linearly combining $\hat D_0$ and $\hat D_1$. As a case study, we will take the following form of linear combination with real valued coefficient $\lambda \in (0,1]$,
\begin{equation}
    \hat D_{\lambda} \equiv (1-\lambda) \hat D_{1} + \lambda \hat  D_{0},\quad \hat H_\lambda \equiv \hat D^\dag_\lambda \hat D_\lambda, \label{def_lambda_model}
\end{equation}
and study the spectrum and geometric properties of the resulting Hamiltonian $\hat H_\lambda$. As discussed above, the $\hat H_\lambda$ is fully gapped in the entire range of $\lambda\in(0,1]$, which is also seen in Fig.~\ref{fig: mix_and_truncated} (a). By varying $\lambda$, one can trace the evolution between the lattice first LL state ($\lambda{=}0$) and lattice LLL states ($\lambda{=}1$). The $\varphi_0$ saturates the trace bound~\cite{JieWang_exactlldescription,Jie_Origin22,kahlerband1,kahlerband2,kahlerband3,Grisha_TBG2,LedwithVishwanathParker22,Emil_constantBerry,Martin_PositionMomentumDuality,RahulRoy14}, but $\varphi_{n>1}$ does not and their integrated trace of metric is not quantized; Schmidt-Gram orthogonalizing $\varphi_{n}$ yields lattice versions of generalized LLs with quantized integrated trace of quantum metric~\cite{GLL_JieWang_2024}. The effect of the truncation to the band gap, band width, quantum geometry and Chern number is also included in Fig.~\ref{fig: mix_and_truncated}. Our model provides guidance in designing large family of tight-binding models exhibiting flatbands with desired band wavefunction and quantum geometries.

\emph{Possible many-body physics, optical lattices and outlooks.---}
The fast decay of hopping amplitude makes the cold-atom realization of the model possible. For ultracold atomic gases in an optical lattice, various techniques to engineer Chern bands have been developed~\cite{jaksch:2003,cooper:2019,agrawal:2024}, which have led to the realizations of complex hopping amplitudes through modulation assisted tunnelings~\cite{aidelsburger:2013,miyake:2013} and complex next-nearest-neighbor hopping through Floquet methods~\cite{jotzu:2014,goldman:2014}.

Including interactions on top of the Chern band is possible to give rise to fractional quantum Hall phases. Pure multi-particle onsite interaction in lattice LLL will be ideal for bosonic Pfaffian state, as it is the parent many-body Hamiltonian. Although high-body interaction is possible to be generated from multi-particle loss processes~\cite{3bodyloss_09,3bodyint_22}, the most natural inter-particle interactions in ultracold atomic gases are two-body on-site Hubbard interaction~\cite{RMP_Bloch_08}. Additionally, numerical studies of interacting bosons with long-range interaction in KM model indicates a wide class of non-Abelian states~\cite{zhao_nonabelian_13}. This also suggests it is promising for having non-Abelian states from short-range interacting bosons in our models. We leave further exploration of interacting physics in our models for future work.

We notice bosonic Pfaffian state from on-site Hubbard interaction has been numerically observed in a Hofstadter model of $\phi=1/6$~\cite{BonsonicPfaffianHofstadter21}. Extending our models from half flux to general flux is possible, and is an interesting future direction. This could offer analytical insight for the origin of the non-Abelian state and help identify wider parameter spaces for these exotic phases, particularly from geometric perspectives.

It is worthy to study the role of higher crystalline symmetries and their interplay with magnetic translation in constraining gapless points. Additionally, it is also possible to extend our model to non-Bravais lattice, where such generalization was already achieved for KM model~\cite{Dong_Mueller_20}. The interaction-driven instabilities in our quadratic band touching models $\hat H_{2n+1}$ is also interesting to explore further~\cite{KaiSun_Quadratic_Touching09}. Last but not least, since the modern theory defines coherent state with respect to a general Lie group~\cite{RMP_coherentstate, perelomov_generalized_1977}, extending the Poisson summation constitutes an interesting mathematical question potentially useful for flatband models and simulating many-body physics.

\emph{Acknowledgement.---} We acknowledge Erich J. Mueller, Nathan Goldman, Zhao Liu, Yinghai Wu, Fengcheng Wu, Manato Fujimoto and Junkai Dong for useful discussions.

G.~J., D.~P. and J.~W. are supported by Temple University start up funding. B.~M. acknowledges support from the Security and Quantum Information Group (SQIG) in Instituto de Telecomunica\c{c}\~{o}es, Lisbon. This work is funded by FCT (Funda\c{c}\~{a}o para a Ci\^{e}ncia e a Tecnologia) through national funds FCT I.P. and, when eligible, by COMPETE 2020 FEDER funds, under Award UIDB/50008/2020 and the Scientific Employment Stimulus --- Individual Call (CEEC Individual) --- 2022.05522.CEECIND/CP1716/CT0001, with DOI 10.54499/ 2022.05522.CEECIND/CP1716/CT0001. T.~O. acknowledges suport from JSPS KAKENHI Grant Number JP24K00548, JST PRESTO Grant No. JPMJPR2353, JST CREST Grant Number JPMJCR19T1. T.~O. and J.~W. acknowledge PCTS, Princeton University for hospitality during the workshop ``New Twist of Quantum Geometry'', during which this work was partially conducted.

Author contributions: X. Shen derived the generalized Poisson sum rule and constructed the theoretical model; G. Ji identified singular points and linearly combined gapped parent Hamiltonians; X. Shen, G. Ji and J. Zhang contributed to the numerical exploration of the model properties; D. Palomino, B. Mera and T. Ozawa contributed to the theory of singular points; J. Wang conceived and supervised the project.

\bibliography{ref.bib}

\appendix

\section*{--- Appendix ---}
In the appendix, we provide more detailed discussions on: coherent states and sum rules, lattice LL states, and model Hamiltonians.

\section*{Coherent states and sum rules}
\subsection{Conventions}
As discussed in the main text, we consider a 2D lattice $\Lambda$ embedded in 2D continuum. We denote primitive vectors as $\bm a_{1,2}$, which encloses area $\bm a_1 \times \bm a_2 = \epsilon_{ab}\bm a_1^a\bm a_2^b = 2\pi l^2$. Here $a,b = x,y$ denote spacial direction and $\epsilon_{xy} = -\epsilon_{yx} = 1$ is the anti-symmetric tensor.

A generic point in the 2D continuum is denoted as $\bm r = r^1\bm a_1 + r^2\bm a_2 = (r^x, r^y)$, where $r^{1,2}\in\mathbb{R}$. A lattice point has integer valued decomposition $m,n\in\mathbb{Z}$ on primitive vectors $\bm r = m\bm a_1 + n\bm a_2 \in \Lambda$. We consider infinity large lattice, so $m,n$ are unbounded.

The mapping from the 2D space to complex plane is determined by the complex structure $z = w_a \bm r^a$. The complex structure $w_{a=x,y} = (1,\tau)/\sqrt{2\tau_2}$ where $\tau = \tau_1 + i\tau_2$ is the moduli parameter. Here $\tau_{1,2}$ are real valued and $\tau_2 > 0$. The complex structure is constrained by $\epsilon^{ab}\bar{\omega}_a\omega_b=i$ and induces a natural uni-modular metric,
\begin{equation}
    g_{ab} = \frac{1}{\tau_2}\left(\begin{matrix}1&\tau_1\\\tau_1&|\tau|^2\end{matrix}\right) = \bar{w}_a w_b + w_a\bar{w}_b,
\end{equation}
that defines distance for the 2D space $g_{ab}\bm r^a\bm r^b = 2z\bar z$.

\subsection{Coherent states}
The Hilbert space of coherent states are specified by the complex structure and the magnetic length $l_B$. The LLL coherent states $\Phi_{0}^{l_B,g}(\bm r)$ referred in this work are defined as states annihilated by the annihilation operator of Landau orbital, and and are expressed as holomorphic functions up to a Gaussian term,
\begin{equation}
    \hat a_{l_B,g} = z/(2l_B) + l_B\bar\partial_z,\qquad \hat a_{l_B,g}\Phi_0^{l_B,g} = 0,
\end{equation}
\begin{eqnarray}
    \Phi_0^{l_B,g}(\bm r) &=& f(z)\exp\left(-\frac{|z|^2}{2l_B^2}\right),\\
    &=& f(z)\exp\left(-\frac{\bm r^2}{4l_B^2}\right),\quad z = w_a\bm r^a,\nonumber
\end{eqnarray}
where $\bm r$ is a point in the 2D continuum. Here $\bm r^2$ stands for $g_{ab}\bm r^a\bm r^b$.The subscripts and superscripts $l_B, g$ emphasize they are determined by the characteristic length $l_B$ and metric $g_{ab}=\bar{\omega}_a \omega_b+\omega_a \bar{\omega}_b$ (equivalently, the complex structure). The $\Phi_0^{l_B,g}$ is the LLL state of a Galilean invariant LL model $\hat H_{\rm LL} = g^{ab}\bm\pi_a\bm\pi_b$ where $\bm\pi_a = -i\bm\partial_a - ie\bm A_a$ are the dynamical momentum obeying the algebra $[\pi_a, \pi_b] = i\epsilon_{ab}l_B^{-2}$.

\subsubsection{Squeezed coherent states}
As is illustrated above, the creation/annihilation operator of Landau orbital, $\hat a^\dag_{l_B,g} = \bar z/(2l_B) - l_B\partial_z$ and $\hat a_{l_B,g} = z/(2l_B) + l_B\bar\partial_z$, are defined respect to a given metric (complex structure). This representation of the ladder operator has action on the Gaussian factor. These operators with different metrics are related by an unitary (Bogoliubov) transformation generated by the squeeze operator,
\begin{eqnarray}
    \hat a_{l_B,g'}&=&\hat{S}^{\dagger}_{l_B,g}(\zeta)\hat a_{l_B,g}\hat{S}_{l_B,g}(\zeta), \\
    &=&\cosh{r}\ \hat a_{l_B,g}-e^{i\theta}\sinh{r}\ \hat a_{l_B,g}^{\dagger}, \quad \zeta=r e^{i\theta}, \nonumber \\
    \hat a_{l_B,g'}&=&\hat{S}_{l_B,g}^{\dagger}(\zeta)\hat a_{l_B,g}\hat{S}_{l_B,g}(\zeta), \\
    &=&-e^{-i\theta}\sinh{r}\ \hat a_{l_B,g}+\cosh{r}\ \hat a_{l_B,g}^{\dagger}, \quad \zeta=r e^{i\theta}, \nonumber
\end{eqnarray}
\begin{equation}
    \hat{S}_{l_B,g}(\zeta) = \exp\left(\frac{\bar{\zeta}}{2} \hat a_{l_B,g}^2 - \frac{\zeta}{2} \hat a_{l_B,g}^{\dag2}\right) = \hat{S}^\dag_{l_B,g}(-\zeta).
\end{equation}

The metrics and complex structures are related as follow,
\begin{eqnarray}
    \omega'_a &=& \cosh{r}\ \omega_a-e^{i\theta}\sinh{r}\ \bar{\omega}_a, \\
    \bar{\omega}'_b &=& -e^{-i\theta}\sinh{r}\ \omega_b+\cosh{r}\ \bar{\omega}_b,
\end{eqnarray}
and,
\begin{equation}
    g_{ab}'=\bar{w}_a' w_b' + w_a'\bar{w}_b'
\end{equation}
The squeezing operation amounts to a determinant preserving deformation of the metric of the coherent state with $\epsilon^{ab}\bar{\omega}'_a\omega'_b=\epsilon^{ab}\omega_a\omega_b=i $. When considering embedding a 2D lattice into a complex plane, it effectively tunes the shape of the lattice cell on the complex plane while preserving its area. The kernel of $\hat a_{l_B,g'}$, {\it i.e.} states $\Phi_0^{l_B,g'}$, are referred to as the squeezed coherent states of $\Phi_0^{l_B,g}$.

\subsection{Lattice sum rules}
Perelomov introduced a sum rule known as the Perelomov identity (or Poisson summation) for coherent states~\cite{perelomov_completeness_1971}. Considering coherent states of $l_B = l$, they are exactly annihilated by the following lattice summation,
\begin{equation}
    \sum_{\bm r\in\Lambda} \eta_{\bm r} \Phi_0^{l,g}(\bm r) = 0,\quad \forall g,\label{app_perelomov_id}
\end{equation}
where $\eta_{\bm r} = (-1)^{m+n+mn}$ for lattice point $\bm r = m\bm a_1 + n\bm a_2 \in \Lambda$ is the gauge function. It is worthy to emphasize Eqn.~(\ref{app_perelomov_id}) is valid for any choice of the uni-modular metric $g$.

There are few descends of the Perelomov identity Eqn.~(\ref{app_perelomov_id}), which we discuss in below.

\subsubsection{Lattice sum rule for lowest LL states}
The sum rule useful for constructing Kapit Mueller model is the following, which follows directly from Eqn.~(\ref{app_perelomov_id}),
\begin{eqnarray}
    0 &=& \sum_{\bm r \in \Lambda} \eta_{\bm r} W_0^{g}(\bm r) \Phi_0^{l_B,g}(\bm r)\label{app_LLL_sumrule},\\
    W_0^{g}(\bm r) &=& \exp\left[-\frac{\bm r^2}{4}(l^{-2} - l_B^{-2})\right],\quad l_B > l,\nonumber
\end{eqnarray}
where the gauge function $\eta_{\bm r} = (-1)^{m+n+mn}$ is determined by the lattice $\Lambda$. Since coherent states are identical to LLL states, the above identity is equivalent to stating, in the physics content that, any LLL states with $l_B\geq l$ can be exactly annihilated by summing over lattice points weighted by $\eta_{\bm r}W_0^g(\bm r)$,
\begin{equation}
    \sum_{\bm r \in \Lambda} \eta_{\bm r} W_0^{g}(\bm r) \Phi_0^{l_B,g}(\bm r) = 0.
\end{equation}

When $l = l_B$, the above sum rule reduces to Eqn.~(\ref{app_perelomov_id}). However, for the purpose of utilizing the sum rule to construct parent Hamiltonians with local hopping, the condition $l_B > l$ is needed. We leave discussions of the Kapit Mueller model Hamiltonian in the following section. 


\subsubsection{Lattice sum rule for even-indexed LL states}
We now discuss generalization of the Perelomov identity to all even-indexed LL states. We start from the initial identity Eqn.~(\ref{app_perelomov_id}) and treat coherent states equipped with an arbitrary metric $\tilde{g}$ as squeezed coherent states of another metric $g$.
\begin{eqnarray}
    \sum_{\bm r\in\Lambda} \eta_{\bm r}\langle\bm r|\Phi^{l,\tilde{g}}_{0}\rangle &=& \sum_{\bm r\in\Lambda} \eta_{\bm r}\langle\bm r|\hat {S}_{l,g}(\zeta)|\Phi^{l,g}_{0}\rangle \\
    &=&\sum_{\bm r \in \Lambda}\sum_{m= 0}^{\infty}\eta_{\bm r}\langle\bm r|\chi_{2m}(\zeta)|\Phi^{l,g}_{2m}\rangle =0  \quad \zeta\in \mathbb{C} \nonumber
\end{eqnarray}
The $\chi_{2n}(\zeta)$ function takes the  form,
\begin{equation}
    \chi_{2n}(\zeta)=\frac{\sqrt{(2 n)!}}{2^{n} n!}\frac{\left(- \tanh r\right)^{n}}{\sqrt{\cosh r}}  e^{i n\phi}, \quad \zeta=r e^{i\phi}
\end{equation}
This series of functions satisfies the following orthogonality relation,
\begin{gather}
     \int_{\mathbb{C}} \bar{\chi}_{2n}(\zeta)\chi_{2m} (\zeta)\mathrm{d}^2\zeta=N_{n}\delta_{nm} \label{app_chi_proj} \\
     N_{n}=\int_{0}^{\infty} \frac{(\tanh{r})^{2n}}{\cosh{r}} \mathrm{d} r
\end{gather}
We utilize Eqn.~(\ref{app_chi_proj}) as a projector to obtain,
\begin{gather}
    N_{n}\sum_{\bm r \in \Lambda}\sum_{m= 0}^{\infty}\eta_{\bm r}\int_{\mathbb{C}}\bar{\chi}_{2n}(\zeta)\chi_{2m}(\zeta)\mathrm{d}^2\zeta\langle\bm r|\Phi^{l,g}_{2m}\rangle =0, 
\end{gather}

This leads to,
\begin{equation}
    \sum_{\bm r \in \Lambda} \eta_{\bm r} \Phi_{2n}^{l,g}(\bm r) = 0.\label{app_zak_id}
\end{equation}

Eqn.~(\ref{app_zak_id}) is the generalized Perelomov identity for all even LL states and is initially derived by M.Boon and J.Zak through a different method~\cite{Boon_Zak_81,Boon_Zak_83}. Notably, this sum rule is not valid for odd LL states.

\subsubsection{Lattice sum rules for all LL states}
We now utilize the generalized Perelomov identity to construct lattice sum rules for all LL states. Define the creation/annihilation operator of Landau orbital that effectively acts on the LL wavefunction up to the Gaussian term. For lowest LL wavefunction, it's the holomorphic subspace $f(z)$. 
\begin{equation}
    \hat{a}_{l_B,f}=l_B\bar{\partial}_z,\quad \hat{a}^{\dagger}_{l_B,f}=\bar{z}/l_B-l_B\partial_z.
\end{equation}

We take $\Phi_{2n}^l=[\hat{a}^{\dagger 2n}_{l_B,f}f(z)]e^{-|z|^2/2l^2}$, and consider general $n \in \mathbb{N}$. We first rewrite the even LL sum rule as follows,
\begin{equation}
    \sum_{\bm r\in\Lambda} \eta_{\bm r} [(\bar{z}-l^2\partial_z)^{2n}f(z)]e^{-|z|^2/{2}l^{2}}=0,
\end{equation}
which is equivalent to,
\begin{equation}
    \sum_{\bm r\in\Lambda} \eta_{\bm r} \{[\bar{z}(\bar{z}-2l^2\partial_z)-l^4{\partial}_z^2 ]^n f(z)\}\cdot e^{-|z|^2/{2}l^{2}}=0.
\end{equation}
It has the binomial expansion,
\begin{eqnarray}
     &&\sum_{\bm r\in\Lambda}\sum_{m=0}^{n}\eta_{\bm r}\binom{n}{m} \{[\bar{z}(\bar{z}-2l^2\partial_z)]^m(-l^4{\partial}_z^2)^{n-m}  f(z)\} \nonumber \\ 
     &&\cdot e^{-|z|^2/{2}l^{2}} 
     =0, \qquad \forall n \in \mathbb{N},
\end{eqnarray}
as $(-l^4\partial_z^2)$ keeps the holomorphic subspace intact, and it applies to all natural number $n$, one find each monomial of the polynomial yields a lattice zero summation,
\begin{equation}
    \sum_{\bm r\in\Lambda} \eta_{\bm r}\{[\bar{z}(\bar{z}-2l^2\partial_z)]^n  f(z)\}\cdot e^{-|z|^2/{2}l^{2}}=0,   \quad  \forall n \in \mathbb{N},
\end{equation}
if we take $l_B^2=2l^2$, and intentionally split the Gaussian term into two parts with one being the Gaussian pocket for LL wavefuntion and the other being part of the weight for lattice summation, we obtain,
\begin{equation}
    \sum_{\bm r\in\Lambda} \eta_{\bm r} \bar{z}^ne^{-|z|^2/{2}l_B^{2}}[a^{\dagger n}_{l_B}  f(z)]\cdot e^{-|z|^2/{2}l_B^{2}}=0,
\end{equation}
which is equivalent as,
\begin{equation}
    \sum_{\bm d \in \Lambda}\eta_{\bm d}\bar{d}^n e^{-|z|^2/{2}l_B^{2}}\Phi^{l_B}_{n}(\bm d)=0, \quad l_B^2=2l^2,
\end{equation}
valid for all $n$ with $l_B^2 = 2l^2$. The above equation is then established as a lattice sum rule with local weight for all LLs.

\section*{Lattice LL States}
The lattice LL states are defined as the continuum LL states sampled on lattice. They are given by,
\begin{equation}
    |\varphi_{n\bm k}\rangle \equiv \sum_{\bm r \in \Lambda} \Phi_{n\bm k}^{l_B}(\bm r)|\bm r\rangle,\label{deflatticeLL}
\end{equation}
where $|\bm r\rangle$ is a complete and orthogonal local real space basis states.

The state $\varphi_{n\bm k}$ is un-normalized. If $\varphi_{n\bm k}$ is defined, the normalized state will be $N_{n\bm k}\varphi_{n\bm k}$ where the normalization factor,
\begin{equation}
    N^{-2}_{n\bm k} = \sum_{\bm r \in \Lambda} |\Phi_{n\bm k}^{l_B}(\bm r)|^2,
\end{equation}
and in general has $\bm k-$dependence. If $\varphi_{n\bm k}$ vanishes at some momentum points $\bm k_*$, the lattice LL states are not well defined at $\bm k_*$.

Here we briefly comment on the quantum geometries of the lattice LL states. It is also worthy to notice that $\varphi_{0\bm k}$ is an ideal Kähler band (lowest generalized LL), a subject discussed in Ref.~(\onlinecite{JieWang_exactlldescription,GLL_JieWang_2024,kahlerband1,kahlerband2,kahlerband3,LedwithVishwanathParker22,Grisha_TBG2}). The $\varphi_{0\bm k}$ saturates the trace bound $\int d^2\bm k ~ {\rm Tr}g_{\bm k} = |C|$, following the fact that its periodic part $u_{0\bm k}$, defined as $e^{-i\bm k\cdot\bm r}\varphi_{0\bm k}$ is a holomorphic function of momentum $\bm k$. For general momentum points, the higher LL lattice states are not orthogonal to each other, because the orthogonality of LL states requires continuous integration, but here the norm is defined respect to lattice summation, $\sum_{\bm r \in \Lambda} N_{m\bm k}\Phi_{m\bm k}^*(\bm r)N_{n\bm k}\Phi_{n\bm k}(\bm r) \neq \delta_{mn}$. Therefore, $\phi_{n>0,\bm k}$ does not qualify to be the generalized LL states proposed in Ref.~(\onlinecite{GLL_JieWang_2024}), so their integrated trace of quantum metric are not expected to take quantized values. They are rather lattice version of the modulated LL states~\cite{GLL_JieWang_2024}. However, if consecutive Schmidt-Gram orthogonalization is performed, generalized LL states are generated and quantized integrated trace of quantum metric is expected.

\subsection{Translation symmetry}
\subsubsection{Continuum LL states: wavefunction and translational properties}
The continuum LL wavefunctions $\Phi_{n\bm k}^{l_B}(\bm r)$ are quasi-periodic when translated on magnetic unit cell. Since this work mainly focus on half flux lattice, without loss of generality, one can choose the magnetic primitive lattice vectors to be $\tilde{\bm a}_1 = 2\bm a_1$ and $\tilde{\bm a}_2 = \bm a_2$. The momentum $\bm k$ labeling the wavefunction is spanned by $\tilde{\bm b}^1 = \bm b^1/2$ and $\tilde{\bm b}^2 = \bm b^2$.

The quasi-periodicity of the LL wavefunctions is,
\begin{equation}
    \Phi_{n\bm k}^{l_B}(\bm r + \tilde{\bm a}_i) = -e^{{{i}\tilde{\bm a}_i\times\bm r}/2l_B^2}e^{i\bm k\cdot\tilde{\bm a}_i}\Phi_{n\bm k}^{l_B}(\bm r),\quad i=1,2.\label{app_mag_trans}
\end{equation}

Concrete form of the LLL wavefunction on torus can be obtained by the modified Weierstrass sigma function $\sigma(z)$~\cite{haldaneholomorphic,Jie_MonteCarlo},
\begin{equation}
    \Phi_{0\bm k}^{l_B}(\bm r) = \sigma(z-z_k)e^{\bar z_kz/l_B^2}e^{-(|z|^2+|z_k|^2)/{2l_B^2}},
\end{equation}
where $z_k = -ik$ and the quasi-periodic domain of $\sigma(z)$ is $\tilde{\bm a}_{1,2}$. The $\Phi_{0\bm k}(\bm r)$ is the momentum resolved LLL state, hence obviously it belongs to $\Phi_0{(\bm r)}$. Higher LL wavefunctions are obtained by applying ladder operators. For instance, the first LL wavefunction is,
\begin{equation}
    \Phi_{1\bm k}(\bm r) = \left[(\bar z - \bar z_k) - \zeta(z-z_k)\right]\Phi_{0\bm k}(\bm r),
\end{equation}
where $\zeta(z) = \sigma'(z)/\sigma(z)$ is the modified Weierstrass zeta function.

\subsubsection{Lattice translation operator}
A lattice magnetic translation $\hat t(\bm d)$ is defined to hop particle by lattice vector $\bm d = m\bm a_1 + n\bm a_2$. The translation operator has to obey the magnetic translation algebra to take into account the nonzero flux,
\begin{equation}
    \hat t(\bm d) \hat t(\bm d') = e^{-{i}\bm d\times\bm d'/{2l_B^2}}\hat t(\bm d + \bm d).\label{magtrans}
\end{equation}

One choice of the concrete form of $\hat t(\bm d)$ is given as follows,
\begin{equation}
    \hat t(\bm d) = \sum_{\bm r \in \Lambda} e^{{i}\bm d\times\bm r/{2l_B^2}} |\bm r-\bm d\rangle\langle\bm r|.\label{app_td_sym}
\end{equation}
Alternative forms can be obtained by applying a site-dependent unitary transformation. For instance, for $\bm r = m\bm a_1 + n\bm a_2$, denoting $\bm r_1 = m\bm a_1$ and $\bm r_2 = n\bm a_2$, the local unitary transformation,
\begin{equation}
    |\bm r\rangle \rightarrow e^{-{i}\bm r_1 \times \bm r_2/{2l_B^2}}|\bm r\rangle = e^{-i \pi m n \phi}|\bm r\rangle,
\end{equation}
maps the translation operator $\hat t(\bm d)$ to,
\begin{equation}
    \hat t(\bm d) \rightarrow e^{-{i}\bm d_1\times\bm d_2/{2l_B^2}}\sum_{\bm r \in \Lambda}e^{i\bm d_1\times\bm r_2/l_B^2}|\bm r-\bm d\rangle\langle\bm r|,\label{app_td_landau}
\end{equation}
where $\bm d_{1,2}$ stands for the first and second components of lattice vector: $\bm d_1 = m\bm a_1$ and $\bm d_2 = n\bm a_2$.

We will term Eqn.~(\ref{app_td_sym}) and Eqn.~(\ref{app_td_landau}) as translation operator in symmetric gauge and Landau gauge, respectively. The Landau gauge is only used for numerical studies. We will proceed the analytical derivation with symmetric gauge.

For following derivations, it is useful to notice the action of translation operator on lattice states,
\begin{equation}
    \hat t(\bm d)|\varphi_{n\bm k}\rangle = \sum_{\bm r \in \Lambda}e^{\frac{i}{2}\bm k\cdot\bm r}\Phi_{n,\bm k-\bm k_{\bm r}}(\bm d)|\bm r\rangle,\label{app_trans_assistant}
\end{equation}
where $(\bm k_{\bm r})_a \equiv -\epsilon_{ab}\bm r^b/l_B^2$. In deriving the above, the following identity is useful, which follows from the magnetic translation properties of continuum LL states,
\begin{equation}
    \Phi_{n\bm k}(\bm r + \bm d) = \Phi_{n,\bm k-\bm k_{\bm r}}(\bm d)e^{\frac{i}{2}\bm k\cdot\bm r}e^{-{i}\bm d\times\bm r/{2l_B^2}}. \label{app_mag_trans_k}
\end{equation}

Eqn.~(\ref{app_trans_assistant}) will later be used to prove the zero mode of the Kapit-Mueller model (for $n=0$) and its generalization studied in this work ($n>0$).

\subsection{Spinor representation}
In this section, we focus on half flux lattice. Since there are two lattice points in each magnetic unit cell, the lattice LL state can be represented as a two component spinor. The summation of lattice points in constructing the lattice LL state can be separated into two parts,
\begin{eqnarray}
    |\varphi_{n\bm k}\rangle &=& \sum_{m,n\in\mathbb{Z}} \Phi_{n\bm k}(2m\bm a_1+n\bm a_2)|2m\bm a_1+n\bm a_2\rangle,\nonumber\\
    &+& \sum_{m,n\in\mathbb{Z}} \Phi_{n\bm k}(\bm a_1+2m\bm a_1+n\bm a_2)|\bm a_1+2m\bm a_1+n\bm a_2\rangle,\nonumber\\
    &=& \sum_{\tilde{\bm a}} \Phi_{n\bm k}(\tilde{\bm a})|\tilde{\bm a}\rangle + \Phi_{n\bm k}(\bm a_1+\tilde{\bm a})|\bm a_1+\tilde{\bm a}\rangle,
\end{eqnarray}
where we have taken the magnetic unit cell to be spanned by $2\bm a_1$ and $\bm a_2$. The $\tilde{\bm a}$ denotes the origin of the magnetic unit cell. After using the magnetic translation properties of the continuum state Eqn.~(\ref{app_mag_trans}), one arrives at,
\begin{eqnarray}
    |\varphi_{n\bm k}\rangle
    &=& \Phi_{n\bm k}(\bm 0)|R_{\bm k}\rangle + \Phi_{n\bm k}(\bm a_1)|R'_{\bm k}(\bm a_1)\rangle,
\end{eqnarray}
where orthogonal states $|R_{\bm k}\rangle = \sum_{\tilde{\bm a}}\eta_{\tilde{\bm a}}e^{i\bm k\cdot{\tilde{\bm a}}}|\tilde{\bm a}\rangle$ and $|R'_{\bm k}\rangle = \sum_{\tilde{\bm a}}\eta_{\tilde{\bm a}}e^{\frac{i}{2}\tilde{\bm a}\times{\bm a_1}}e^{i\bm k\cdot\tilde{\bm a}}|\tilde{\bm a}+{\bm a_1}\rangle$.
The two independent components $\Phi_{n\bm{k}}(\bm{0})$ and $\Phi_{n\bm{k}}(\bm{a}_{1})$
span a spinor.

Meanwhile, the LL wavefunction also obeys the following inversion symmetry:
\begin{equation}
    \Phi_{n,\bm k}(\bm r) = (-1)^{n+1}\Phi_{n,-\bm k}(-\bm r). \label{app_inversion_sym}
\end{equation}
By combining the inversion  and translation properties Eqn.~(\ref{app_mag_trans}), (\ref{app_mag_trans_k}), (\ref{app_inversion_sym}) of LL wavefunctions, one can derive that the lattice wave function $|\varphi_{n\bm k}\rangle$ vanish at $\bm{k}{=}\tilde{\bm{b}}_{1}/2$ and $(\tilde{\bm{b}}_{1}+\tilde{\bm{b}}_{2})/2$, i.e.,
\begin{align}
    \Phi_{2n+1,\tilde{\bm{b}}_{1}/2}(\bm{0})&=\Phi_{2n+1,\tilde{\bm{b}}_{1}/2}(\bm{a}_{1})=0, \\
    \Phi_{2n+1,(\tilde{\bm{b}}_{1}+\tilde{\bm{b}}_{2})/2}(\bm{0})&=\Phi_{2n+1,(\tilde{\bm{b}}_{1}+\tilde{\bm{b}}_{2})/2}(\bm{a}_{1})=0.
\end{align}
It leads to that the energy bands of $H_{2n+1}$'s are gapless at these points. These gapless points are singular points studied in Ref.~\cite{rhim:yang:19} and the corresponding flat bands are singular flat bands. Note that there is a trivial momentum shift between the symmetric gauge and Landau gauge, due to gauge transformations. In Section ``zero mode operator'', we comment how to see the gapless point complementarily from the symmetry properties of the zero-mode operator.

\subsection{Projector and topology of the odd lattice LL states}
In this part, we discuss the mathematical aspects of the singular points in more detail.

For odd $n$, the projector $P_{n\bm{k}}:=\frac{|\varphi_{n\bm k}\rangle \langle\varphi_{n\bm k}|}{\sqrt{\langle\varphi_{n\bm k}|\varphi_{n\bm k}\rangle}}$ is well-defined for all $\bm{k}$ except the $\bm{k^*}$ singular points. This is a non-trivial result despite the fact that $|\varphi_{n\bm k}\rangle$ vanishes at those singular points. This is because, in many cases, $|\varphi_{n\bm k}\rangle$ = 0 at $\bm{k} = \bm{k^*}$ does \textit{not} imply $\lim_{\bm{k}\rightarrow \bm{k^*}}P_{n\bm{k}}$ does not exist. In the case that the limit does exist, we can define a \textit{new} projector $\tilde{P}_{n\bm{k}}:=P_{n\bm{k}}$ for $\bm{k} \neq \bm{k^*}$ and $\tilde{P}_{n\bm{k^*}}:=\lim_{\bm{k}\rightarrow \bm{k^*}}P_{n\bm{k}}$. This gives a new well-defined projector that is continuous over the entire Brillouin Zone where we can recover our band topology. As stated above, this limit does not exist for odd n.

Another way to understand this is that there are singularities in Bloch wavefunctions which are removable by a gauge transformation and those are not. A simple example is that of a linear band crossing in two dimensions. In that case, assuming the crossing occurs at $\bm{k}=0$, the Bloch Hamiltonian, near that point, can be modeled by $H=k_x\sigma_x +k_y\sigma_y$, where $\sigma_x$ and $\sigma_y$ are Pauli matrices. The unnormalized Bloch wavefunction for the upper band can be chosen to be $|u_{\bm{k}}\rangle= (|z|, z)$, for $z=k_x+ik_y$. Observe that the wavefunction vanishes at the band crossing point. This singularity is not removable. Indeed, we can choose the representative $(1,z/|z|)$ for the Bloch wavefunction and we see that we can not extend the wavefunction to the origin. The corresponding orthogonal projector is $P(\bm{k})=\frac{1}{2}\begin{bmatrix}
1 & \bar{z}/|z|\\    
z/|z| & 1 \end{bmatrix}$, which is not defined at the origin. Removable singularities occur generically when one has a rank $1$ globally defined periodic orthogonal projector $P(\bf{k})$ and hence a well-defined Bloch (line) bundle, but the former is not topologically trivial. In this case, one cannot hope to find a globally defined periodic Bloch wavefunction, which forces us to work with locally defined nonvanishing Bloch wavefunctions (i.e., local sections of the Bloch bundle) or with globally defined nonvanishing but multivalued Bloch wavefunctions. In the former case, if we try to extend the local section to a global section we will necessarily find zeros (or other kinds of singularities), while in the latter case the singularity arrises in the multivaluedness of the Bloch wavefunction. 

Now let us focus on the topological classification, from the Bloch line bundle point of view, in the case there are singularities. Suppose we have $n$ singularities, with $n\geq 1$. In our case $n=2$, but it is instructive to keep the discussion general. Then, we can define a Bloch line bundle over the Brillouin zone with $n$ punctures, i.e., with $n$ points removed. The Brillouin zone with $n$ puctures is a topological space that has the homotopy type of an $n$-bouquet, which topologically is the same as the wedge sum of $n$ circles. This is a $1$-dimensional space and cannot support second cohomology, where the first Chern class naturally lives. Since the first Chern class classifies line bundles up to isomorphism (in the smooth category), the Bloch line bundle is trivializable in this case.




\section*{Model Hamiltonians}
\subsection{Zero mode operator}
The zero mode operator introduced in the main text,
\begin{equation}
    \hat D_n = \sum_{\bm d \in \Lambda} W_n(\bm d) \hat t(\bm d).
\end{equation}

In this section, we will discuss in detail why such operator exactly annihilates lattice LL states. In addition, we will discuss symmetry constrains on such operators, and the enforced vanishing points.

\subsubsection{Zero energy mode}
Here we prove the following properties on lattice of $\phi = l^2/l_B^2$ number of flux quanta per unit cell.
\begin{itemize}
    \item For any $\phi \in (0,1)$, the $\hat D_0$ has zero modes $\varphi_{0\bm k}$.
    \item On half-flux lattice $\phi = 1/2$, the $\hat D_n$ of all $n$ has right zero modes $\hat D_n|\varphi_{n\bm k}\rangle = 0$.
    \item For any $\phi \in (0,1)$, the $\hat D^\dag_n$ of all $n$ has right zero modes $\hat D^\dag_n|\varphi_{0\bm k}\rangle = 0$.
\end{itemize}

The first two statements follows from the lattice sum rule discussed above. To see this, we apply $\hat D_n$ to $\varphi_{n\bm k}$ and use Eqn.~(\ref{app_trans_assistant}),
\begin{eqnarray}
    \langle\bm r|\hat D_n |\varphi_{n\bm k}\rangle &=& \sum_{\bm d,\bm r\in\Lambda} W_n(\bm d) \langle\bm r|\hat t(\bm d)|\varphi_{n\bm k}\rangle,\\
    &=& \sum_{\bm r\in\Lambda} e^{\frac{i}{2}\bm k\cdot\bm r} \left[\sum_{\bm d\in\Lambda}W_n(\bm d)\Phi_{n,\bm k-\bm k_{\bm r}}(\bm d)\right],\nonumber
\end{eqnarray}
which yields zero after applying generalized sum rules. The last statement follows directly from the LLL lattice sum rule. We notice $\hat D_0$ is precisely the Kapit-Mueller model (when adding an extra on-site chemical potential term to the original model)~\cite{Kapit_Mueller}. The $\hat H_n = \hat D^\dag_n\hat D_n$ is the new model studied in this work whose zero-energy flatband is $\varphi_{n\bm k}$.

\subsubsection{Symmetry properties and gapless points}
We denote the inversion operator as $\hat I$, which is a unitary operator that maps local basis $|\bm r\rangle$ to its inversion partner $\hat I|\bm r\rangle = |-\bm r\rangle$. The inversion transformation maps the translation operator to its inverse,
\begin{equation}
    \hat I\hat t(\bm d)\hat I^{-1} = \hat t(-\bm d),
\end{equation}
and consequently, the zero mode operator transforms as,
\begin{equation}
    \hat I\hat D_n\hat I^{-1} = (-1)^n \hat D_n.
\end{equation}
In other words, even indexed operator commute with inversion $[\hat D_{2n}, \hat I] = 0$ and the odd indexed operator anti-commute with inversion $\{\hat D_{2n+1}, \hat I\} = 0$.

See see the gapless points from the zero-mode operator, we seek the momentum representation of $\hat D_n$ in the Landau gauge Eqn.~(\ref{app_td_landau}). In such gauge, the translation is generated by two elementary translations,
\begin{eqnarray}
    \hat t_1 &\equiv& \hat t(\bm a_1) = \sum_{\bm r\in\Lambda} e^{i\bm a_1\times\bm r_2/l_B^2}|\bm r-\bm a_1\rangle\langle\bm r|,\\
    \hat t_2 &\equiv& \hat t(\bm a_2) = \sum_{\bm r\in\Lambda} |\bm r-\bm a_2\rangle\langle\bm r|,
\end{eqnarray}
where it is easy to verify $\hat t_1\hat t_2 = -\hat t_2\hat t_1$. We will expand the translation operator and the zero-mode operator in the momentum basis $|\bm k\rangle$ defined as $|\bm k\rangle = \sum_{\bm r\in\Lambda}\exp(i\bm k\cdot\bm r)|\bm r\rangle$. Here $\bm k$ belongs to the un-folded Brillouin zone spanned by $\bm b_{1,2}$; as we will soon see, the magnetic translation will down-fold the Brillouin zone to a reduced zone with half area. The actions of the elementary translations to momentum basis are given by,
\begin{eqnarray}
    \hat t_1|\bm k\rangle &=& e^{2\pi i k_1}|\bm k+\bm b_2/2\rangle,\\
    \hat t_2|\bm k\rangle &=& e^{2\pi i k_2}|\bm k\rangle,
\end{eqnarray}
where $\bm k = k_1\bm b_1 + k_2\bm b_2$. The momentum point in the reduced zone will be defined as the simultaneous eigenstate of $\hat t_1^2$ and $\hat t_2$,
\begin{eqnarray}
    \hat t_1^2|\tilde{\bm k}\rangle &=& e^{2\pi i \tilde{k}_1} |\tilde{\bm k}\rangle,\\
    \hat t_2|\tilde{\bm k}\rangle &=& e^{2\pi i \tilde{k}_2} |\tilde{\bm k}\rangle,
\end{eqnarray}
where $\bm k = \tilde{k}_1 \tilde{\bm b}_1 + \tilde{k}_2 \tilde{\bm b}_2 = k_1\bm b_1 + k_2\bm b_2$ where $\tilde{\bm b}_1=\bm b_1/2$ and $\tilde{\bm b}_2=\bm b_2$. The $\tilde k_1 = 2k_1 \in [0,1)$ and $\tilde k_2 = k_2 \in [0,1)$.

The matrix element of the zero-mode operator, at a point in the reduced Brillouin zone, is,
\begin{equation}
    \hat D_{n,\bm k} = \left(\begin{matrix} \langle\bm k|\hat D_n|\bm k\rangle & \langle\bm k|\hat D_n|\bm k+\bm b_2/2\rangle \\ \langle\bm k+\bm b_2/2|\hat D_n|\bm k\rangle & \langle\bm k+\bm b_2/2|\hat D_n|\bm k+\bm b_2/2\rangle \end{matrix}\right).
\end{equation}
The symmetry properties $\hat I\hat D_n = (-1)^n \hat D_n\hat I$, $\hat I|\bm k\rangle = |-\bm k\rangle$, $|\bm k+\bm b\rangle = |\bm k\rangle$ implies the following identities,
\begin{equation}
    \hat D_{n,\bm k} = (-1)^n \hat D_{n,-\bm k},\quad \hat D_{n,\bm k} = \hat D_{n,\bm k+\bm b}.
\end{equation}
Hence we proved the $\hat D_{n,\bm k}$ becomes a two by two zero matrix at inversion symmetric points $\bm 0$, $\bm b_{1,2}/2$ or $(\bm b_1+\bm b_2)/2$ when the LL index is $2n+1$. Since the reduced Brillouin zone is half of the Brillouin zone, it only contains two of the gapless points $\bm 0$ and $\bm b_2/2$.

\subsection{Explicit Tight-binding Form}
In this section, we provide explicit form of the tight-binding model. Since our model's hopping amplitude has Gaussian decay, one can truncate the model parameters while well approximately preserving key features, such as narrow band and quantum geometries.

In the exact limit, the model Hamiltonian $\hat H_n$ reads,
\begin{eqnarray}
    \hat H_n &=& \hat D^\dag_n \hat D_n,\\
    &=& \sum_{\bm d,\bm d'\in\Lambda} W^*_n(-\bm d')W_n(\bm d) \hat t(\bm d') \hat t(\bm d),\nonumber\\
    &=& \sum_{\bm d,\bm d'\in\Lambda} e^{-\frac{i}{2l_B^2}\bm d'\times\bm d}W^*_n(-\bm d')W_n(\bm d) \hat t(\bm d+\bm d'),\nonumber
\end{eqnarray}
where magnetic translation algebra is used.

To proceed, we introduce $\bm d_+ \equiv \bm d+\bm d'$ and $\bm d_- \equiv \bm d-\bm d'$, which are lattice points. However, notice $\bm d_\pm$ are not independent: the $\eta_{\bm d_+}$ is locked to be identical to $\eta_{\bm d_-}$.
By using $\eta_{\bm d}\eta_{\bm d'} = \eta_{\bm d_+}\exp(\pm i\bm d_+\times\bm d_-/4l^2)$, one arrive at the compressed form of the model Hamiltonian which is one of the key result derived in this section,
\begin{equation}
    \hat H_n = \sum_{\bm d\in\Lambda} J_n(\bm d) \hat t(\bm d),\label{app_tb_model}\\
\end{equation}
where the tight-binding coupling is given by,
\begin{eqnarray}
    J_n(\bm d) &=& \eta_{\bm d} G_n(\bm d) e^{-(l^{-2} - l_B^{-2})\bm d^2/4},\\
    G_n(\bm d_+) &=& \sum_{\bm d_-\in\bm d_++2\Lambda} w^n(\bm d_+,\bm d_-) e^{-(l^{-2} - l_B^{-2})w^*(\bm d_+,\bm d_-)},\nonumber
\end{eqnarray}
where the function $w(\bm d_+, \bm d_-) = (d_- - d_+)(\bar d_- + \bar d_+)/4$. In the above, un-bolded letter $d_\pm$ refers to the complex coordinate of $\bm d_\pm$. The summation over $\bm d_+ + 2\Lambda$ means summing over lattice points $(m_+ + 2m)\bm a_1 + (n_+ + 2n)\bm a_2$ where $\bm d_+ = m_+\bm a_1 + n_+\bm a_2$. The half-flux condition $l_B = \sqrt{2} l$ should be imposed.

\end{document}